\documentclass[aps,prl,twocolumn,superscriptaddress,nofootinbib]{revtex4}

\usepackage{graphicx}
\usepackage{amssymb}
\usepackage{amsmath}

\input h+a.mac

\begin{document}
%\preprint{MSUHEP-yymmdd}

%\title{Charged and neutral Higgs boson associate  production at hadron
%     colliders: the test of the light Higgs scenario}
\title{Light MSSM Higgs boson scenario and its test at hadron colliders}

\author{Alexander Belyaev}
\affiliation{Department of Physics and Astronomy,
          Michigan State University,
      East Lansing, MI 48824, USA}
\author{Qing-Hong Cao}
\affiliation{Department of Physics,
      University of California at
      Riverside, CA 92521,
      USA}
\author{Daisuke Nomura}
\affiliation{Department of Physics and Astronomy,
          Michigan State University,
      East Lansing, MI 48824, USA}
\author{Kazuhiro Tobe}
\affiliation{Department of Physics and Astronomy,
          Michigan State University,
      East Lansing, MI 48824, USA}
\author{C.-P. Yuan}
\affiliation{Department of Physics and Astronomy,
          Michigan State University,
      East Lansing, MI 48824, USA}

%\date{\today}

\begin{abstract}
We show that in the Minimal Supersymmetric Standard Model,
the possibility for the lightest CP-even Higgs boson  to be
lighter than $Z$ boson (as low as about 60 GeV) is, 
contrary to the usual belief, not yet
excluded by LEP2 data or any other existing experimental data. The
characteristic of the light Higgs boson scenario (LHS) is that the $ZZh$
coupling and the decay branching ratio ${\rm Br}(h/A\to b\bar{b})$
are simultaneously suppressed as a result of generic
supersymmetric loop corrections. Consequently, the $W^\pm H^\mp h$
coupling has to be large due to the sum rule of Higgs couplings to
weak gauge bosons. In addition to discussing the potential of the Tevatron
and $B$-factories to test the LHS, we show that
the associate neutral and  charged Higgs boson
production process, $pp\to H^\pm h (A)$, can completely
probe LHS at the CERN Large Hadron Collider.
%\vspace*{-0.5cm}
%\vskip -1cm
\end{abstract}
%                             % Classification Scheme.
\maketitle
%\tableofcontents

%\section{Introduction}
%
While the Standard Model (SM) of
particle physics is consistent with existing data, 
there is a strong belief to a more complete
description of the underlying physics. 
Supersymmetry (SUSY), as a good candidate for theory beyond the SM, 
solves principal theoretical problems of the SM such as hierarchy and fine
tuning, as well as provides good dark matter candidate 
and potentially solves the problem of baryogenesis. 
In the minimal 
supersymmetric standard model (MSSM)~(for example, see \cite{Haber:1984rc}),
the Higgs sector consists of {\it two} doublet fields $h_d$ and $h_u$ 
to generate masses for down- and up-type fermions, respectively, 
and to provide an anomaly-free theory.
After spontaneous symmetry breaking, there remain
five physical Higgs bosons: a pair of charged
Higgs bosons $H^{\pm}$, two neutral CP-even scalars 
$H$ (heavier) and $h$ (lighter), and a neutral
CP-odd pseudoscalar $A$. 
Higgs potential is constrained
by supersymmetry such that all the tree-level Higgs boson
masses and self-couplings are determined by only two independent unknown
parameters, commonly chosen to be the mass of the CP-odd
pseudoscalar ($M_A$) and the ratio of  vacuum expectation 
values  of neutral Higgs fields, 
denoted as $\tan \beta\equiv \langle h_u\rangle /\langle h_d\rangle$.

The MSSM predicts a light neutral Higgs boson which is 
lighter than $Z$-boson at the tree level,
since the Higgs quartic coupling is determined by the SM gauge couplings.
However, large top quark and squark~(stop) loop contributions induce  significant
radiative correction to the Higgs quartic coupling, such that the lighter neutral Higgs
boson mass can be as large as 
$130$ GeV~\cite{Okada:1990vk,Haber:1990aw,Ellis:1990nz,Barbieri:1990ja}
and avoid the LEP2 limit.
The negative result of Higgs boson search at LEP2 
via $e^+e^-\to Z h$ production channel imposes a lower 
bound on the SM Higgs boson mass $M_h >114~{\rm GeV}$~\cite{Barate:2003sz}.
This limit can be translated into constraint on the Higgs sector of MSSM, 
which excludes significant
portion of SUSY parameter space.

The LEP2 collaborations 
have performed analyses for the MSSM~\cite{unknown:2006cr} 
using several benchmark scenarios
that  were considered as typical cases for MSSM parameter space. In
this Letter, we propose a different branch of the MSSM parameter space 
which has not been previously studied  with deserved attention. 
We call this possibility 
light Higgs boson scenario (LHS),
in which the lightest Higgs boson is lighter than the $Z$-boson  
and generic MSSM  radiative corrections induce  significantly small $ZZh$ coupling so that the LEP2
constraint from $e^+e^-\to Z h$ production channel can be avoided. 
The similar possibility was previously noted in Ref.~\cite{Kane:2004tk} but
without detailed study of  MSSM parameter space. 
Here we consider only the MSSM without
CP-violation (for CP-violating case, see Ref.~\cite{Carena:2000ks})
and
specify the generic MSSM parameter space
for LHS scenario to be consistent with the LEP2 and other existing experimental
constraints.
We discuss the potential of the  Fermilab Tevatron and $B$-factories
to test the LHS and show that the CERN Large Hadron Collider (LHC)
can completely probe LHS via the neutral and charged Higgs
boson associate production process $pp\to H^\pm h /H^\pm A$.

LEP2 collaborations analyzed especially two complementarity processes
for MSSM Higgs boson search: 
$e^+e^-\to Zh/Ah$~\cite{unknown:2006cr}, in which
 the first one occurs via $ZZh$ coupling $g_{ZZh}
=\sin(\beta-\alpha)(\equiv s_{\beta\alpha})$ while the second one  via
$ZAh$ coupling $g_{ZAh} = \cos(\beta-\alpha)$. The
obvious sum rule ($g_{ZZh}^2+g_{ZAh}^2=1$) puts strong constraints
on the mass and couplings of the MSSM Higgs boson $h$.
For all studied  benchmark scenarios at LEP2, $M_h$ below about
$90$ GeV is excluded~\cite{unknown:2006cr}. In this study we
demonstrate that LEP2 has missed the generic parameter space,
outside of the benchmark points, with  $60 \, \mbox{GeV}\lsim M_h
\lsim M_Z$.

In order to satisfy the LEP2 constraint derived from the
production channel $e^+e^- \rightarrow Zh$ with $M_h<M_Z$, the
coupling $g_{ZZh}$ ({\it i.e.} $s_{\beta\alpha}$) has to be small.
Here, we describe in detail the mechanism for suppressing the
$g_{ZZh}$ coupling.
Let us denote ${\cal M}^2$ for a $2\times 2$ squared-mass matrix of
the CP-even neutral Higgs bosons in the gauge eigenbasis $({\rm
Re}~h_d^0,{\rm Re}~h_u^0)$.
The mass eigenstates $(h, H)$  are given by the diagonalization of
the matrix ${\cal M}^2$ with the definition:
\begin{equation}
\label{hmix}
\left(
\begin{matrix}
   h \\
   H
\end{matrix}
\right)
=
 \left(
  \begin{matrix}
   -\sa & \ca \\
    \ca &  \sa
 \end{matrix}
\right)
\left(
\begin{matrix}
 {\rm Re}~ h_d^0\\
 {\rm Re}~ h_u^0
\end{matrix}
\right),
\end{equation}
where  $c_\alpha\equiv\cos\alpha~{\rm and}~s_\alpha
\equiv\sin\alpha$ (where $-\pi/2\leq\alpha\leq\pi/2$). Using the components
of  matrix ${\cal M}^2_{ij}$, $s_{\beta\alpha}$
can be analytically
expressed as
\begin{equation}
s_{\beta\alpha}=
\frac{(D+x)^{1/2} s_\beta \pm (D-x)^{1/2} c_\beta}
{\sqrt{2D}},
\end{equation}
where $s_\beta\equiv \sin\beta$, $c_\beta\equiv \cos\beta$,
$x\equiv {\cal M}^2_{11}-{\cal M}^2_{22}$, $y\equiv{\cal
M}^2_{12}$, $D\equiv \sqrt{x^2+4y^2}$, and the signs ``$\pm$''
correspond to negative and positive $y$, respectively. As to be
shown below, the LHS requires $\tan\beta>1$ to be consistent with
 experimental data.
For relatively large $\tan\beta$  ($s_\beta \gg c_\beta$)
and $y/x\simeq 0$, we obtain
$s_{\beta\alpha}\simeq \frac{(|x|+x)^{1/2}}{\sqrt{2|x|}}=0 $
which takes place for $x<0$. Therefore, conditions $y/x\simeq 0$
and $x<0$ provide small values of  $s_{\beta\alpha}$. We note that
when $M_A>M_Z$ and $\tan\beta>1$, $x=(M_A^2-M_Z^2)(-\cos2\beta)>0$
and $y=-\frac{M_A^2+M_Z^2}{2}\sin2\beta$ at tree level.
Therefore, the loop corrections to $x$ and $y$ are very important
and have to be as large as the tree-level values in order to
satisfy the conditions $y/x\simeq 0$ and $x<0$, which yield a small
value of $s_{\beta\alpha}$. It is important  to mention, however,
that tree-level values of $x$ and $y$ are naturally suppressed
when $M_A\simeq M_Z$ and $\tan\beta$ is large.
When $y/x\simeq 0$ and $x<0$, the lightest neutral Higgs boson $h$
mainly consists of $h_d^0$, and the neutral Higgs boson masses are
approximately given by $M^2_h\simeq {\cal M}^2_{11}$ and
$M^2_H\simeq {\cal M}^2_{22}$, which is different from the usual
scenarios. As it is well-known, the ${\cal M}^2_{22}$ 
(i.e.~$h_u$-component) receives
large positive logarithmic correction from top and stop
contributions,
$\delta {\cal M}^2_{22}\simeq \frac{3 y_t^4 v^2 s_\beta^2}{8
\pi^2} \ln \left(\frac{M_S^2}{m_t^2}\right)$\,,
where $y_t$ is the top Yukawa coupling and $M_S$ is the average
stop
mass~\cite{Okada:1990vk,Haber:1990aw,Ellis:1990nz,Barbieri:1990ja}.
This correction, which helps to significantly increase the mass of
$h$ in the usual scenarios, increases the mass of $H$ in the LHS
case. Therefore, even though $x>0$ at tree level, the
condition $x<0$ can be easily realized in the LHS because the large
logarithmic correction to ${\cal M}^2_{22}$ at the one-loop level
can overcome  its small tree-level value  when $M_A\sim M_Z$. In order to
realize the condition $y/x\sim 0$, certain values of the trilinear A-term 
of the stop ($A_3$) and the supersymmetric Higgs mass $\mu$-parameter ($\mu$) 
are usually required, depending on $M_A$,
$\tan\beta$ and other  SUSY breaking parameters.
Typically, $|A_3|>400$ GeV and
$\mu>300$ GeV, as we will show later.

\begin{table}
\begin{tabular}{|l|l|l|l|l|l|l|l|}
\hline
%point #92
$\tan\beta$ & $M_{H^+}$     & $\mu$     & $A_3$     &  $M_1$ &   $M_2$    & $M_3$     & $M_{Q}$\\
$35$        & $130$         & $700$ &   $700$       &  $150$ &   $300$    & $600$     & $320$  \\
\hline
\hline
%\multicolumn{7}{|l|}{$M_h=77$, $M_A=104$, $M_H=120$} \\
%\multicolumn{7}{|l|}{${\rm Br}(h/A/H\to b\bar{b})= 0.70/0.68/0.45$} \\
%\multicolumn{7}{|l|}{${\rm Br}(h/A/H\to \tau\bar{\tau})= 0.30/0.31/0.45$} \\
%\multicolumn{7}{|l|}{$g_{ZZh}^2=0.017,~ g_{ZZH}^2=0.983$} \\
%\multicolumn{7}{|l|}{$M_{\tilde\chi^+_1}=295$, $M_{\tilde{t}_1}=125$, $M_{\tilde{b}_1}=270$} \\
%\multicolumn{7}{|l|}
%{$\Delta\rho=6.1\times 10^{-4} $}\\
\multicolumn{8}{|l|}{$M_h=79$, $M_A=104$, $M_H=119$} \\
\multicolumn{8}{|l|}{${\rm Br}(h/A/H\to b\bar{b})= 0.69/0.68/0.47$} \\
\multicolumn{8}{|l|}{${\rm Br}(h/A/H\to \tau\bar{\tau})= 0.30/0.31/0.45$} \\
\multicolumn{8}{|l|}{$g_{ZZh}^2=0.019,~ g_{ZZH}^2=g_{H^+ W^- h}^2=0.981$} \\
\multicolumn{8}{|l|}{$M_{\tilde\chi^+_1}=295$, $M_{\tilde{t}_1}=129$, $M_{\tilde{b}_1}=270$} \\
\multicolumn{8}{|l|}
{$\Delta\rho=5.8\times 10^{-4} $}\\
\hline
\end{tabular}
\caption{\label{tab:bench}LHS sample point for MSSM parameters at the weak  scale. The
dimension of mass parameters is in units of GeV. 
$M_i(i=1,\ldots, 3)$, $M_{Q}$ and $A_3$ are gaugino masses, the universal soft-breaking sfermion mass
and universal trilinear A-term for the third-generation at the weak scale, respectively.
$M_{\tilde\chi^+_1}$, $M_{\tilde{t}_1}$ and $M_{\tilde{b}_1}$ are pole masses
for the lightest chargino, stop and sbottom, respectively.
}
\end{table}
We present one of our sample points for the LHS in Table~\ref{tab:bench}. 
For simplicity, we assume $M_2=2 M_1$ for gaugino masses, 
the universal soft-breaking sfermion mass ($M_Q$) and trilinear A-term ($A_3$)
for the third-generation at the weak scale. For our numerical
analysis, we use CPsuperH program~\cite{Lee:2003nt} and assume CP
is conserved.
Although at  tree level, for the sample point, $x>0$,
$y/x\simeq -0.22$ and $s_{\beta\alpha}\simeq 0.98$, the Higgs mass 
matrix elements in the effective potential 
become ${\cal M}^2_{11}\simeq (82.0~{\rm GeV})^2,~{\cal
M}^2_{22}\simeq (120~{\rm GeV})^2, ~{\cal M}_{12}^2\simeq
-(29.1~{\rm GeV})^2$, and hence $x<0$ and $y/x\simeq 0.11$, after
including the radiative corrections 
(we use 172.5 GeV top-quark mass~\cite{Group:2006hz} in our studies).
Consequently, we can obtain
small $s_{\beta \alpha}$ ($s_{\beta\alpha}\simeq 0.14$) since both
the conditions $x<0$ and $y/x\simeq 0$ are realized by including
the large radiative corrections. Note that for the LHS the mass term of $h_d$-component
${\cal M}^2_{11}$ does not receive as large radiative corrections
as ${\cal M}^2_{22}$ does, and hence the lighter Higgs mass
is close to its tree-level value
$M_h\simeq \sqrt{{\cal M}^2_{11}}\sim M_Z$ when $M_A\sim M_Z$.
This feature 
is qualitatively very different from those in the commonly
discussed MSSM scenarios. 
On the contrary, the mass of the heavier CP-even
Higgs boson receives large radiative corrections
to exceed about 114 GeV in order to be in agreement with  LEP2 data,
since the  $ZZH$ coupling is close to the SM value.

To search for the LHS parameter space,
we scan the following set of MSSM parameters: $\tan\beta~[1.1, 50]$, 
$(M_{1}/{\rm TeV})~[0.05, 1]$, $(M_3/{\rm TeV})~[0.05, 1]$, 
$(A_3/{\rm TeV})~[-2, 2]$, $(M_{Q}/{\rm TeV})~[0.05, 1]$ and 
$(\mu/{\rm TeV})~[0, 3M_Q]$, within the range denoted in brackets.
Since a too large $\mu$-parameter induces not only the color breaking vacuum in the general direction
of the scalar potential but also the fine-tuning in the Higgs mass parameter,
we require  $\mu$ to be less than $3M_Q$ in our analysis~\cite{Drees:2005jg}.
Then, we check the LHS parameter space against the full set of the experimental 
and theoretical constraints.
 The relevant constraints are the following:
(1) LEP2 $Zh/ZH$ and $Ah/AH$ constraints~\cite{unknown:2006cr};
(2) Chargino ($M_{\tilde{\chi}^+_1}$), stop ($M_{\tilde{t}_1}$), sbottom ($M_{\tilde{b}_1}$) and gluino ($M_3$) 
  mass limits: 
%chargino
$M_{\tilde{\chi}^+_1}>103$~GeV~\cite{Yao:2006px},
%stop
$M_{\tilde{t}_1}>96$~GeV~\cite{Yao:2006px},
%sbottom
$M_{\tilde{b}_1} > 220$ GeV for $M_{\tilde{\chi}^0_1}<90$ GeV
and $M_{\tilde{b}_1} - M_{\tilde{\chi}^0_1} > 6$~GeV (where $M_{\tilde{\chi}^0_1}$ is the 
neutralino mass)~\cite{Abazov:2006fe}
or $M_{\tilde{b}_1} >100$ GeV for all other regions~\cite{Yao:2006px},
and
%gluino
$M_3 > 270$~GeV for $M_{\tilde{b}_1} < 220$~GeV
and $M_3 - M_{\tilde{b}_1} > 6$~GeV~\cite{Abulencia:2005us} or
$M_3>240$~ GeV for all other regions~\cite{Abazov:2006bj};
(3) electroweak constraint: one-loop stop
contributions to $\rho$-parameter $|\Delta\rho_{\rm stop}| < 2\times
10^{-3}$~\cite{Drees:1990dx};
(4) color breaking constraint:  
$A_3^2<3(2M_{Q}^2+M_{h_u}^2+\mu^2)$ where $M_{h_u}$ is the soft-breaking mass 
for Higgs $h_u$~\cite{Frere:1983ag,Claudson:1983et}.

\begin{figure}[htbp]
\includegraphics[width=0.46\textwidth]{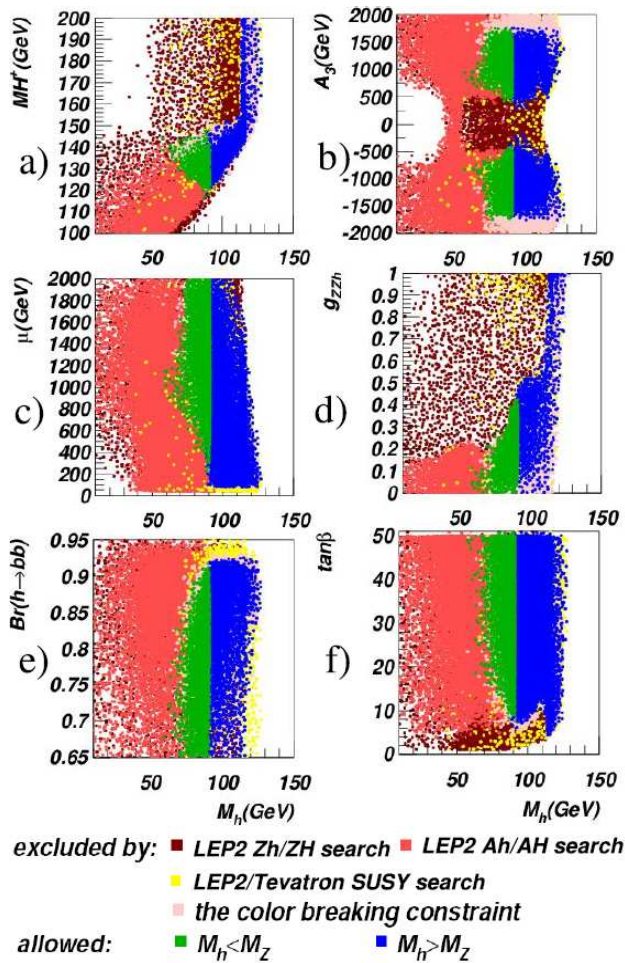}
\caption{\label{2d}
Projected planes of scanned parameter space indicating LHS region in accord with
experimental data.
See detail explanation in the text.
}
\end{figure}

The results shown in Fig.~\ref{2d} unveil that essential LHS parameter space
survives all constraints and the mass of the light Higgs boson can be as low as
about 60~GeV. In Fig.~\ref{2d}, green (blue) color indicates allowed parameter space
with $M_h<M_Z$ ($M_h>M_Z$).
All other colors indicate regions excluded by
LEP2 $Zh/ZH$ search~(dark red),
LEP2 $Ah/AH$ search~(red),
direct LEP2/Tevatron SUSY searches~(yellow)
and color breaking constraint~(light red).
%and ${\rm Br}(b\to s\gamma)$ constraint (gray).
We note that $\Delta\rho$ constraint
does not further limit the parameter space
once LEP2 Higgs boson search and
SUSY particle search constraints are applied.
Fig.~\ref{2d}a ($M_{H^+}$-$M_{h}$ plane) shows that LHS scenario
(green) is realized for low values of charged Higgs boson mass:
$120~\mbox{GeV}<M_{H^+}<150$~GeV, indicating the non-decoupling
regime. Much lighter charged Higgses
are excluded mainly by the LEP $Ah$ production constraint.
The scenario requires intermediate-to-large values of the A-term and
$\mu$-parameter; 
$|A_3|>400$~GeV and $\mu\gsim 300$GeV (cf. Fig.~\ref{2d}b(c):
$A_3(\mu)$-$M_{h}$ plane) to make $g_{ZZh}$ small, as indicated in
Fig.~\ref{2d}d.
On the other hand, larger {\it positive} \ $M_3\mu\tan\beta$ \ product 
gives rise to larger
{\it negative}  correction to the bottom Yukawa coupling $y_{hbb}$.
This large negative correction to $y_{hbb}$ is {\it non-universal}
with respect to the $\tau$ Yukawa coupling $y_{h\tau\tau}$ and
leads to a suppression in ${\rm Br}(h/A \to
b\bar{b})$~\cite{Carena:1998gk}
large enough to avoid LEP2 constraint from the $Ah$ channel 
with low $M_h$ values. This channel is complementary to
the LEP2 $Zh$ search in excluding light Higgs bosons since
$g_{ZAh}$ coupling is enhanced when $g_{ZZh}$ is suppressed. 
In the LHS parameter space ${\rm Br}(h/A \to b\bar{b})$ can
be suppressed down to about 50\% (cf. Fig.~\ref{2d}e), and
consequently ${\rm Br}(h/A \to \tau\bar{\tau})$ is enhanced up
to about 50\%, so that  $Ah$ channel is not observed:
$bbbb$ decay mode is largely suppressed, while $bb\tau\tau$ or
$\tau\tau\tau\tau$ signatures are not enhanced enough to exclude
60 GeV$\lesssim M_h < M_Z$.
Fig.~\ref{2d}e
presents the ${\rm Br}(h\to b\bar{b})$--$M_h$ correlations.
It is interesting to note that the relatively large $\mu$-parameter simultaneously
suppress both $s_{\beta\alpha}$ and ${\rm Br}(h/A \to b\bar{b})$ to be consistent with the LEP2
constraints.
We also note that as $\tan\beta$ gets larger, the lighter Higgs
becomes possible (Fig.~\ref{2d}f).
It is worth mentioning that although the heavy Higgs $H$ couplings to vector bosons are SM-like,
its couplings to down-type fermions are further suppressed as compared to those of
light and CP-odd Higgs bosons (see Table~\ref{tab:bench}).

Since in LHS $g_{ZZh} (=s_{\beta\alpha})$ is suppressed, $H^+ W^-
h$ coupling is inevitably enhanced due to the sum rules in Higgs
boson couplings to weak gauge bosons, i.e., $g_{ZZh}^2+g_{H^+ W^-
h}^2=1=g_{H^+ W^- A}^2$. Therefore,  
$q\bar{q}'\to H^\pm h (A)$ production via $W$
boson exchange could be sizable with the production cross section
$\sim$ 10 fb at the Tevatron and $\sim$ 100 fb at the
LHC for $M_{h/A}\sim 100$ GeV ~\cite{Kanemura:2001hz,Cao:2003tr}.
\begin{figure}[htb]
\includegraphics[width=0.47\textwidth]{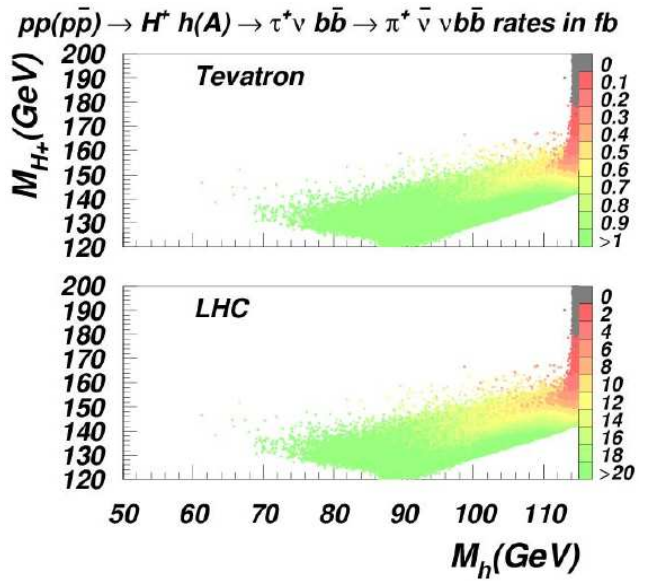}
\caption{\label{fig:cs} Rates for $p{\bar p},\, pp
\to H^+ h(A)\to\tau^+\nu b\bar{b}\to \pi^+\bar{\nu}\nu b\bar{b}$
signature at the Tevatron and the LHC.} 
\end{figure}
In Fig.~\ref{fig:cs} we present the inclusive cross section of the
$p{\bar p},\, pp \to H^+ h(A)\to\tau^+\nu b\bar{b}\to
\pi^+\bar{\nu}\nu b\bar{b}$ signature at the Tevatron and the
LHC in the $M_{H^\pm}$-$M_h$ plane. For simplicity, we have
combined the $H^+ h$ and $H^+ A$ production rates. As clearly
shown in Fig.~\ref{fig:cs}, the LHC can be sensitive to the entire
LHS parameter space, assuming that the above signal event
signature can be measured at the 1~fb level~\cite{Cao:2003tr}.
 The potential of the Tevatron to observe
$H^+ A/H^+ h$ process deserves special investigation and will be
reported elsewhere~\cite{ahp-future}. We also note that when
$s_{\beta\alpha}$ is small, the tree level bottom and $\tau$
Yukawa couplings are enhanced by a factor of
$(-\sin\alpha/\cos\beta)\simeq \tan\beta$, compared with the SM
values. Therefore, the LHS, which is realized in
intermediate-to-high $\tan\beta$ region, can be potentially probed
even at the Tevatron via several {\it $\tan\beta$-enhanced}
processes, such as $p\bar{p}\to h(A)$ with $h/A\to\tau\bar{\tau}$
(produced via gluon-gluon fusion process), $p\bar{p}\to b\bar{b}
h(A)$,  as well as $p\bar{p}\to t\bar{t}$ with $t\to H^+ b$. At present
luminosity, those processes are sensitive only to very high values
of $\tan\beta\gtrsim 45-50$, while at $10$ fb$^{-1}$
$\tan\beta\gtrsim 30$ could be
probed~\cite{Carena:2000yx,Belyaev:2002zz}.
Finally, we note that in the LHS, the flavor physics processes
at $B$-factories and Tevatron,
such as $b\rightarrow s \gamma$
~\cite{Degrassi:2000qf,Carena:2000uj}, $B^-\rightarrow \tau^- \bar{\nu}$
~\cite{Ikado:2006un,Browder_talk},
$B_{d,s}\rightarrow \mu^+\mu^-$~\cite{Babu:1999hn}
and  
$B_s-\bar{B}_s$ oscillation measurements~\cite{Abazov:2006dm,unknown:2006mq}
can be 
largely modified due to the
sizable contributions generated by light Higgs bosons,
although the predictions may
strongly depend on the flavor structure of the SUSY breaking terms. 

%\section{Conclusions}
{\bf Conclusions:} We have found that in the MSSM the possibility  for
the lightest CP-even Higgs boson  to be lighter than 
$Z$-boson (as low as about 60 GeV) is,
contrary to the usual belief,  not yet excluded by existing
experiments. The characteristic of the light Higgs boson  scenario is
that the $ZZh$ coupling and the decay branching ratio ${\rm
Br}(h/A\to b\bar{b})$ are simultaneously suppressed as a result of
SUSY loop corrections. 
We would like to note that the region of the MSSM parameters 
used  for explanations of non-conclusive LEP2 excess 
of  $\sim 98$~GeV 'Higgs-like' events~\cite{Barate:2003sz}
studied in the literature (see e.g.~\cite{Drees:2005jg,Nagoya_group})
is the subset of the more generic LHS parameter space we have found
in this paper. Our result would be useful for clarifying the parameter space
responsible for this excess.

The key-test of the light Higgs boson
scenario is the $pp (p\bar{p}) \to H^\pm
h(A)$ production at hadron colliders: if LHS is indeed realized in
nature, then it will be unambiguously discovered or excluded at
the LHC. Meanwhile, this scenario can be tested at the Tevatron
through various production and decay processes with large
$\tan\beta$ enhancement such as 
$p\bar{p}\to h(A)$ with $h/A\to\tau\bar{\tau}$, $p\bar{p}\to b\bar{b}
h(A)$ and  $p\bar{p}\to t\bar{t}$ with $t\to H^+ b$.
Current and future $B$-factories could also provide important tests of
LHS via $b\rightarrow s \gamma$, $B^-\rightarrow \tau^- \bar{\nu}$,
$B_{d,s}\rightarrow \mu^+\mu^-$ 
and $B_s-\bar{B}_s$ oscillation measurements.

{\bf Acknowledgments:}
We thank M.~Drees, G. Kane, N.~Maekawa and C. Wagner for useful discussions.
%A.~B. acknowledges M.~Drees for useful discussions.
%K.~T. thanks N.~Maekawa for useful discussions on their 
%work~\cite{Nagoya_group}
%related to our work.
C.P.Y. and K.T. thank the National Center for
Theoretical Sciences in Taiwan 
for its hospitality, where part of the work was done. 
This work was supported in
part by the US National Science Foundation under award
PHY-0555545 and the US Department
of Energy under Grant No. DE-FG03-94ER40837.

%%%%%%%%%%%%%%%%%%%%%%%%%
%%%%%%%%%%%%%%%%%%%%%%%%%
\bibliography{h+a}
\bibliographystyle{apsrev}

\end{document}